\documentclass[aps,prl,onecolumn,superscriptaddress,groupedaddress]{revtex4}
\usepackage{subfig}
\usepackage{graphicx}  
\usepackage{dcolumn}   
\usepackage{bm}        
\usepackage{amssymb}   
\usepackage{slashed}
\usepackage{graphicx}				
\usepackage{amsmath}
\usepackage{mathtools}
\usepackage{tikz}
\usetikzlibrary{arrows,backgrounds}
\usepackage[all]{xy}
\usepackage{yfonts}
\usepackage{color}
\usepackage{slashed}
\newcommand{\bra}[1]{\ensuremath{\left\langle#1\right|}}
\newcommand{\ket}[1]{\ensuremath{\left|#1\right\rangle}}
\newcommand{\Bracket}[1]{\ensuremath{\left\langle#1\right\rangle}}

\begin{document}
\title{Anomaly cancellation by generalised cohomology}
\author{Andrei T. Patrascu}
\address{University College London, Department of Physics and Astronomy, London, WC1E 6BT, UK}

\begin{abstract}
Supersymmetric states in M-theory are mapped after compactification to perturbatively non-supersymmetric states in type IIA string theory, with the supersymmetric parts being encoded in the non-perturbative section of the string theory. An observer unable to recognise certain topological features of string theory will not detect supersymmetry. Such relativity of symmetry can also be derived in the context of Theorem 3 in ref. [11]. The tool of choice in this context is the universal coefficient theorem linking cohomology theories with coefficients that reveal respectively hide certain topological features. As a consequence of these observations, it is shown that the same theorem is capable of linking perturbative with non-perturbative string theoretical domains. A discussion of inflow anomaly cancellation is also included in the context of universal coefficient theorems.
\end{abstract}
\maketitle
\section{introduction}
M-theory, as a unified theory of physics, makes no distinction between perturbative and non-perturbative states. Indeed $11$-dimensional M-theory describes both perturbative and non-perturbative effects of ten dimensional superstring theory [1,2]. 
It has been shown that 4-dimensional M-theory vacua with $N>0$ supersymmetry appear to have no unbroken supersymmetry from the perspective of perturbative type IIA string theory. The M-theoretical supersymmetry appears as a non-perturbative effect and is encoded in the appearance of non-trivial Ramond-Ramond (RR) charges [3].  Given a spacetime $X$, in order to obtain the low energy effective theory one compactifies string theory on this spacetime. The result will include many $U(1)$ gauge fields. Performing Kaluza-Klein (KK) reduction of the 10-dimensional type IIA string theory or type IIB supergravity, Ramond-Ramond gauge fields will emerge. Such gauge fields will form a vector space which will be dual to a space of harmonic forms in $X$. This implies [4] that the RR charges will take values in the cohomology of our spacetime with real coefficients $H^{*}(X,\mathbb{R})$. After quantising the RR charge, a cohomology with integer coefficients will replace the previous one $H^{*}(X,\mathbb{Z})$ which, further on, will be replaced with $H^{*}(X,\mathbb{Z}/N)$ for large $N$ [4]. On the other side perturbative string theory can only detect zero RR-charges (the non-trivial charges are invisible from the standpoint of perturbative string theory). In non-perturbative string theory we have a sector of the spectrum associated to D-branes wrapped around supersymmetric cycles $W$ in $X$. These will have non-zero charges under RR gauge fields. Such charges can be calculated in terms of the topology of the embedded cycle $f:W\hookrightarrow X$ and the topology of the Chan-Paton bundle $E\rightarrow W$. Moreover, within the inflow mechanism for anomaly cancellation, the charges of the RR bulk fields are induced by the gauge fields and gravitational curvatures. Contributions to these RR charges also come from certain twisted normal bundles. The three phenomena, namely charge induction, inflow anomaly cancellation, and a relation between perturbative and non-perturbative string theoretical domains may have a common origin, related to the universality of the choice of coefficient structures in (co)homology. As M-theory does include both trivial and non-trivial RR charges and does not make a fundamental distinction between perturbative and non-perturbative effects, it is possible that the arbitrariness related to the choice of the coefficient structure in (co)homology is a new fundamental property of M-theory, not studied before.  
\section{Detectability of topological features}
As there is no distinction to be made between perturbative and non-perturbative states from the perspective of M-theory, it is important to identify a unifying viewpoint relating the situation when RR charges are only seen to be zero and the situation when one classifies them either in terms of cohomology with coefficients in $\mathbb{Z}$, $H^{*}(X;\mathbb{Z})$ or within K-theory (which is yet another form of generalised cohomology). Indeed, such a unifying viewpoint can be obtained by analysing universal coefficient theorems in cohomology as well as the way they connect ordinary and generalised cohomology theories [5-7]. Indeed, the existence and calculability of universal coefficient theorems for various generalised cohomology theories is still subject to intensive research in homological algebra and algebraic topology. This article claims no final mathematical construction. However, the various observations it makes regarding the physical properties of RR-charges and their classification are of importance in identifying relations between perturbative and non-perturbative string theory sectors.

In ref. [13] it is derived, and in [3] it is mentioned that the worldsheet action of the $D=10$ type IIA superstring can be obtained by identifying the third worldvolume coordinate with the eleventh spacetime coordinate in the $D=11$ supermembrane. The eleventh coordinate corresponds to a circle. From a cohomological perspective, detecting a circle depends on the choice of a coefficient structure.
From ref. [15] it is known that when twisted cohomology is employed to analyse a circular subspace of a certain topological space, the cohomology completely ignores the parts of the space formed by circles along which the monodromy of the coefficient system is non-trivial. In another sense, the measuring device (expressed mathematically as the coefficient structure) must be added in order for the cohomology to be able to tell us anything about the topology of the space. However, if we decide to employ a coefficient structure that has non-trivial monodromy around certain circular subspaces, those subspaces will not be visible. From the point of view of cohomology with coefficients having non-trivial monodromy when considered around the circular subspace of our manifold, such circular spaces may as well not exist. However, non-trivial RR charges appear and can be calculated in terms of the topology of the embedded supersymmetric cycles $W\hookrightarrow X$ defined above. One could argue that the freedom given by the ability to arbitrarily choose the coefficients is an unnecessary complication. However, (co)homology theory cannot be defined without such coefficients. Indeed, the coefficient structure is included in the very axioms of cohomology theory (Eilenberg Steenrod axioms) and can only be chosen to be trivial, but never eliminated. Therefore, in the best case we can think of the choice of coefficient structures as of a more generalised gauge choice which can reveal certain topological properties while mask others. Therefore we have to accept that taking different choices of coefficients makes mathematical sense. Once we accept this, we must think about what the physical effects of such choices can be. Phenomena appearing to alter significantly when the coefficient structures are altered may not be fundamental. Coefficient choice invariant phenomena however may be the foundation for a unified theory of nature. M-theory appears to have such properties and appears at least up to a certain point to be constructed in a coefficient-covariant way. 
Twisted acyclicity of a circle means that the complement of the tubular neighbourhood of a link looks like a closed manifold to a twisted cohomology, because the boundary being fibered to circles, is invisible in the twisted cohomology [15]. The same remains valid for a set of pairwise transversal, generically immersed closed manifolds of codimension 2 in arbitrary closed manifolds, with the condition that the monodromy around each such manifold is non-trivial. The modified cohomology does not feel the intersection of the submanifolds as a singularity [15]. This is natural if one thinks that the coefficient structure makes the movement around the circle gain a non-trivial lift due to the existence of the coefficient-bound monodromy. In terms of a twisted (co)homology theory there simply is no intersection. It is important to notice that a particular situation where the inflow anomaly cancellation is relevant is for intersection anomalies, arising for example on I-branes. While this is certainly not the only relevant situation, it is worthwhile to observe that lifting the intersection by means of non-trivial coefficients in cohomology has a similar effect as the inflow mechanism. While the inflow mechanism implies a higher dimensional "flow" towards the anomaly, bringing in an "anti-anomaly" that would cancel the original anomaly and make the theory consistent, the structure of coefficients in cohomology adds the "anti-anomaly" by means of a redefinition of a topological measuring device.
The two effects seem to be related and have as origin a common principle. The historical path to dealing with gauge anomalies was quite impressive. For example the first approach was to postulate the existence of a new fermion that would lead to an anti-anomaly term that would precisely cancel the term of the anomaly existing in the original theory. This idea led to various discoveries of new fermions in nature but it finally proved as an insufficient approach. New, more abstract ways in which anomalies can be cancelled have been introduced, the anomaly Inflow representing the idea that a gauge anomaly does not have to be eliminated if it can be compensated by contributions from the higher dimensional space in which the anomalous theory is embedded. We therefore replaced the introduction of new fermions with the introduction of anti-anomalies arising from higher dimensions. In this article I propose a different approach, as mentioned earlier. The gauge anomalies can be in fact cancelled or compensated for by the effects of a new gauge invariance related to the coefficient structures in cohomology. Changing the coefficient structure in cohomology certainly changes the ability of the respective cohomology to detect certain topological features but being an arbitrary choice, it must not be fundamental to nature, and hence situations differing only by the effects of different choices in cohomology must compensate each other out to produce a coefficient choice invariant description of nature. Several gauge anomalies can therefore be compensated by demanding such higher invariance. 
There is more to it. There are several effects that are here in a sense equivalent. The procedure of anomaly cancellation by higher dimensional Inflow being equivalent to a change in cohomology coefficients can lead us to the conclusion that the additional dimensions used for anomaly cancellation are in fact emerging from a gauge principle that is manifest at the level of the different arbitrary choices of coefficients in cohomology. It seems like there are dual views on this problem, and if we take a look at the dimension axiom in cohomology (and abandon it) we may understand why. 
If $P$ is a one point space then by the dimension axiom the (co)homology $H_{n}(P)=0$ for all $n\neq 0$ and $H_{0}(P)$ is the coefficient group. The coefficient group provides us with the structure of the elemental point of a space in cohomology. Eliminating this axiom as is being done in generalised cohomology allows us to introduce arbitrary coefficients, even in higher orders, leading to non-trivial structures associated to our point. Making this modification within the inner structure of our "point" is apparently equivalent to considering an inflow anomaly cancellation via a relation with a higher dimension. This leads us to the conclusion that additional "inner" structure is dual to additional higher dimensions. If one follows this train of thoughts one can imagine that additional dimensions are an emergent outcome originating from the existence of internal topological structure and associated gauge invariance in our elemental points, related to a different, higher form of gauge symmetry. In that sense coefficient gauge symmetry is dual to emergent higher dimensions. 
 If we are about to look at the cobordism between immersed links, we will see from the perspective of twisted cohomology only a compact cobordism between closed manifolds. This allows us (provided we find a consistent relation between twisted cohomology and cohomology with constant coefficients) to analyse manifolds with links of codimension two as if they were single closed manifolds. All this has been shown in [15] which is a purely mathematical paper. However, the interpretation of all these statements in terms of string theory, M-theory, as well as the connection between perturbative and non-perturbative domains is novel and is a part of this research work. 

 In the case in which a D-brane wraps around a cycle of a curved manifold, its normal bundle may twist leading to chiral asymmetry for the theory in its world volume.
 
 The wrapping and intersecting D-branes will be plagued by anomalies which in general do not cancel among themselves and may appear not to be cancelled by the standard inflow mechanism which one may invoke considering the possibility of imbedding the theory in a higher dimensional structure. To understand this difficulty it is important to understand how the inflow mechanism would work in the cases when it is directly applicable. In such fortunate cases, the anomalous theory can be embedded in a higher dimensional theory. The bigger theory has an associated classical action which will have an anomalous variation which is localised at the world volume for our anomalous theory and cancels its anomaly. The term "inflow" therefore originates from the fact that an "anti-anomaly" flows from the higher dimensional theory leading to the cancellation of our lower dimensional anomaly. The cases in which such a method fails appear due to the fact that the anomalies cannot in certain cases be properly factorised. The reason for such a non-factorability is inherently topological. If D-branes are wrapped around non-trivial cycles of a certain compactification manifold, the anomalies appear in the form of non-vanishing variations of the effective action under a local gauge transformation [12]. This is a catastrophic scenario that must be dealt with. Ref. [16, 17, 18] discuss these issues extensively. Ref. [12] recovers the inflow mechanism by arriving again at a proper factorisation by means of the topological result providing us with a relation between the Thom and Euler classes.  
 
When, in the context of M-theory, the eleventh spacetime coordinate corresponds to a $U(1)$ fibre [3] the membrane worldvolume would correspond to a $U(1)$ bundle over the two dimensional worldsheet of the string and not to a direct product. Dimensional reduction of the $11$ dimensional supergravity on a circle leads to type IIA supergravity. In terms of solutions of these theories, this implies that any solution of the form $M_{10}\times S^{1}$ of $11$ dimensional supergravity can be seen as a solution of type IIA superstring theory [3]. This example however restricts the argument to direct product solutions. This however is not required.

 Any solution that has the form of a $U(1)$ bundle over a $10$ dimensional base manifold admits a $10$ dimensional interpretation. An example discussed in [3] is the $AdS_{4}\times S^{7}$ case for $11$ dimensional supergravity, precisely because $S^{7}$ has the form of a $U(1)$ bundle over $CP^{3}$. If the bundle is non-trivial, the KK vector potential that appears after reducing the $11$ dimensional theory to the $10$ dimensional theory becomes topologically non-trivial. It is worthwhile mentioning that every circle on $S^{7}$ can be shrunk to a circle, in contrast to the $CP^{2}\times S^{1}$ case.

In ref. [8] it is noted that a key feature of K-theory is that when comparing two objects, $X'$ and $X''$, it is allowed to augment them by some object $Y$. In condensed matter (the subject of ref. [8]) such augmentation is done by a trivial system, a procedure known in high energy physics from the BRST-anti-BRST quantisation [9]. Two systems may not be trivially deformable one into the other, but still, after such an extension, trivial deformation becomes possible. In cohomology theory such an augmentation is controlled by the coefficient structure. Indeed, in ordinary cohomology theory the cohomology associated to a point is trivial i.e.  $H^{0}(Pt)=\mathbb{Z}$ and $H^{n}(Pt)=0$ for $n>0$, where $Pt$ represents a point space. 
The zero order cohomology of the point $H^{0}(Pt)$ represents the coefficient structure. In generalised cohomology theory however, the point may obtain additional structure, first by allowing the zero order cohomology to become non-trivial, and second, by allowing non-zero higher cohomology groups for the point. For a brief definition of these concepts see [10]. 
Therefore, a broad spectrum of additional structures can be included in a theory only by departing from the trivial cohomology structure. Indeed, I showed in [7] that a global anti-anomaly can be introduced only by means of non-trivial coefficient structure in cohomology. That such a global anti-anomaly can play the same role as an extra-dimensional flow compensating the anomalies by means of the inflow technology can be seen as follows. I will mostly use the notation of [12]. 

\section{Alternative anomaly cancellation}
Let $M$ be the $m$-dimensional world volume of the brane and let $\mathcal{L}_{M}$ be the Lagrangian density controlling the dynamics on the brane. To introduce the brane into the bulk theory one has to add to the bulk action the term 
\begin{equation}
\int_{M}\mathcal{L}_{M}
\end{equation}
If we want to express this in terms of an integral over the whole bulk spacetime $X$ we introduce, following [12], an additional differential form $\tau_{M}$ such that 
\begin{equation}
\int_{M}\zeta = \int_{X}\tau_{M}\wedge \zeta
\end{equation}
for any rank-$m$ form $\zeta$ defined over $M^{*}$. The rank of our additional form is equal to the codimension of $M$ in $X$. This equation defines $\tau_{M}$ as an element of the dual of the space of forms i.e. the space of currents. These are the differential form analogues of distributions and $\tau_{M}$ is the appropriate generalisation of the Dirac delta distribution.

If $\zeta$ is restricted to be a closed form on $M$, the equation above defines only a cohomology class $[\tau_{M}]$, known as the Poincare dual of $M$. It contains topological information about $M$. $\tau_{M}$ itself is a representative of this class. In physics, this is associated to a brane current of the brane wrapped around $M$.

This can be expressed in terms of exact sequences as follows. In the most general case, if we have a subspace of a space then we can think of it as filling out some of the directions in the large space and then, we may define an orthogonal complement that fills out the other directions. Together they span the full space in a minimal way. In a sense, this can be extended to the idea of D-branes in our spacetime. However, when additional structure is added not by looking "outside" at the large vector space, but instead by adding structure to points via coefficients in (co)homology, forming the orthogonal complements may become more complicated because we may not have an immediate inner product (or some suitable pairing) to rely upon. In the most general case, if $A$ is a subspace of $B$ and $A$ fills up certain directions in $B$, the remaining directions are encoded in $B/A$. If $A$ is precisely the kernel of the surjection $B\rightarrow C$ then $A$ fills out some of the directions in $B$ and all the complementary directions are encoded in $C$. This means basically just that $0\rightarrow A\rightarrow B\rightarrow C\rightarrow 0$ is a short exact sequence. 
Therefore, a sequence being exact means we can write 
\begin{equation}
\int_{M}\zeta = \int_{X}\tau_{M}\wedge \zeta
\end{equation}
in a global sense and hence $\tau_{M}$ is globally meaningful. 
As the objects we deal with here (namely $\tau_{M}$ and $\zeta$) are identified as cohomology (resp. homology) classes, what we need to analyse is a pairing between homology and cohomology. This is where universal coefficient theorems enter our discussion.
 In the most general case, consider our spacetime $X$ and our D-branes $M_{1}$ and $M_{2}$ represented algebraically as chain complexes over a ring $R$. Then there exists an evaluation map 
\begin{equation}
Hom_{R}(X,\mathcal{A}_{1})\times X\rightarrow R
\end{equation}
providing us with the evaluation (a pairing)
\begin{equation}
(f,z)\rightarrow f(z)
\end{equation}
Such a pairing passes to the Kronecker pairing 
\begin{equation}
<,>:H^{q}(X;\mathcal{A}_{2})\times H_{q}(X;\mathcal{A}_{1})\rightarrow R
\end{equation}
relating homology with cohomology. Such a pairing is bilinear and its adjoint is a homomorphism 
\begin{equation}
H^{q}(X,\mathcal{A}_{2})\rightarrow Hom(H_{q}(X,\mathcal{A}_{1});\mathcal{A}_{2})
\end{equation}
with coefficients $\mathcal{A}_{1}$ and $\mathcal{A}_{2}$
It however need not be an isomorphism. Universal coefficient theorems provide a measure of how this adjoint fails to be an isomorphism in terms of $Ext$ and $Tor$ groups. The exact sequence 
\begin{widetext}
\begin{equation}
0\rightarrow Ext_{R}(H_{q-1}(X;\mathcal{A}_{2}),A_{1})\rightarrow H^{q}(X;\mathcal{A}_{2})\rightarrow Hom(H^{q}(X;\mathcal{A}_{1}),\mathcal{A}_{2})\rightarrow 0
\end{equation}
\end{widetext}
shows that the $Ext$ group needs to be added in order for the sequence to be exact. When this happens the homology and cohomology with the different coefficients define the integral over the entire space $X$ in a consistent manner. 

In what follows we will see that the global definition of $\tau_{M}$ and the expression $\tau_{M_{1}}\wedge \tau_{M_{2}}$ governing the I-brane are crucial for the cancellation of anomalies by means of inflow techniques. Defining such global structures within integrals covering the whole space $X$ therefore relies on the existence of exact sequences associated to (co)homology theories with coefficients expressed in terms of potentially intersecting D-branes. The correction to $\tau_{M_{1}}\wedge \tau_{M_{2}}$ due to global effects is therefore encoded in the $Ext$ group of the universal coefficient theorem in (co)homology.

In string theory let $M$ be the worldvolume of a $D$-brane. The RR potential called $C$ couples to it and the brane current $\tau_{M}$ arises. The RR field strength is denoted by $H$. On $M$ we have the tangent bundle of the total spacetime $X$ as being $T(X)$ which decomposes into the Whitney sum of the tangent and normal bundles to $M$, namely $T(M)$ and $N(M)$.

Let us start now the other way around. We have the worldvolume of a D-brane and over each point on this D-brane we add the structure of $H^{0}(Pt)$, the zero order cohomology of the point. Then, according to the specific problem, if our D-brane wraps around topologically non-trivial structures of our bulk spacetime $X$, we may add to each point higher cohomology structures $H^{q}(Pt)$. The more complex the topology of the bulk space, the more structure one has to encode in the coefficients of the (co)homology in order to represent the fact that the D-brane can probe it.  

We can construct locally 
\begin{equation}
\tau_{M}=\delta(x^{1})dx^{1}\wedge ... \wedge \delta(x^{dim N(M)})dx^{dim N(M)}
\end{equation}
where $x^{\mu}$ are Gaussian normal coordinates in the transverse space of $M$, or equivalently Cartesian coordinates in the fibre of $N(M)$. Such an expression is however not defined globally. The intersection of two brane world-volumes produces a so called I-brane $M_{12}=M_{1}\cap M_{2}$. Following the assumption of right angles in [12], the tangent bundle of the total spacetime decomposes as 
\begin{widetext}
\begin{equation}
T(X)=T(M_{1})\cap T(M_{2})\oplus T(M_{1})\cap N(M_{2})\oplus N(M_{1})\cap T(M_{2})\oplus N(M_{1})\cap N(M_{2})
\end{equation}
\end{widetext}
The intersection is assumed to be fibre-wise. 
Clearly, 
\begin{equation}
T(M_{12})=T(M_{1})\cap T(M_{2})
\end{equation}
and 
\begin{equation}
N(M_{12})=T(M_{1})\cap N(M_{2})\oplus N(M_{1})\cap T(M_{2})\oplus N(M_{1})\cap N(M_{2})
\end{equation}
Then it follows that $\tau_{M_{1}}\wedge \tau_{M_{2}}=\tau_{M_{12}}$ provided that $N(M_{1})\cap N(M_{2})=\emptyset$. Given the local form of $\tau_{M}$ above, in any other case $\tau_{M_{12}}=0$. 
On the I-brane there can be an anomaly of the form 
\begin{equation}
I_{12}=\pi\int \tau_{M_{12}}\wedge (Y_{1}\wedge\tilde{Y}_{2}+Y_{2}\wedge \tilde{Y}_{1})
\end{equation}
where $Y_{i}$ and $\tilde{Y}_{i}$, $i=1,2$ are some invariant polynomials of the Yang-Mills field strengths and gravitational curvatures on $M_{i}$. With this notation and the results of [12] the anomaly can be canceled if one implements the following Ansatz for Chern-Simons type action on D-branes
\begin{widetext}
\begin{equation}
-\frac{\mu}{2}\sum_{i}\int_{M_{i}}N_{i}C-(-1)^{q}H\wedge Y_{i}^{(0)}=-\frac{\mu}{2}\sum_{i}\int_{X}\tau_{M_{i}}\wedge (N_{i}C-(-1)^{q}H\wedge Y_{i}^{(0)})
\end{equation}
\end{widetext}
In this equation $q$ is $1$ for the II-A string theory and $0$ for the II-B string theory and $i$ labels the $D$-brane wrapping worldvolume $M_{i}$ whose brane current is $\tau_{M}$. $N_{i}$ is the constant part of $Y_{i}$. $C$ and $H$ represent formal sums of all the RR antisymmetric tensor potentials and field strengths respectively. When we integrate we implicitly take products of forms with the required total rank. The rank will appear as an indexation for a formal sum, for example, for a type IIA string theory we have 
\begin{equation}
C=C_{(1)}+C_{(3)}+C_{(5)}+C_{(7)}+C_{(9)}
\end{equation}
Note that $H$ will have corrections to its usual expression $dC$. Given the coupling of the Ansatz above, the equations of motion are 
\begin{equation}
d*H=\mu\sum_{i}\tau_{i}\wedge Y_{i}
\end{equation}
with the Bianchi identities being 
\begin{equation}
dH=-\mu\sum_{i}\tau_{i}\wedge \tilde{Y}_{i}
\end{equation}
where 
\begin{equation}
\tilde{Y}_{j(l)}=-(-1)^{\frac{dim(M_{j})-q}{2}}(-1)^{\frac{l}{2}}Y_{j(l)}
\end{equation}
With these conditions we have 
\begin{equation}
H=dC-\mu(-1)^{q}\tau_{M}\wedge  \tilde{Y}_{j}^{(0)}
\end{equation}
where $\tilde{Y}_{j}^{(0)}$ is the secondary characteristic. We need to observe that the field strengths $H$ are physical observables and therefore must be gauge invariant. Therefore, to any gauge variations, the variations of $C$ must have a compensating nature
\begin{equation}
\delta_{g}C=\mu\sum_{j}\tau_{M_{j}}\wedge \tilde{Y}_{j}^{(1)}
\end{equation}
where $\tilde{Y}_{j}^{(1)}$ is the Wess-Zumino descent of $\tilde{Y}_{j}$. 
The variation of the Ansatz under gauge transformations is 
\begin{equation}
\begin{array}{c}
\delta_{g}S=-\frac{\mu^{2}}{2}\sum_{ij}\int_{X}\tau_{M_{i}}\wedge \tau_{M_{j}}\wedge (\tilde{Y}_{j}^{(1)}N_{i}+\tilde{Y}_{j}(Y_{i})^{(1)})=\\
=-\frac{\mu^{2}}{2}\sum_{ij}\int_{X}\tau_{M_{i}}\wedge \tau_{M_{j}}\wedge (Y_{i}\wedge \tilde{Y}_{j})^{(1)}\\
\end{array}
\end{equation}
This would cancel the anomaly if $\frac{\mu^{2}}{2}=\pi$ when $N(M_{1})\cap N(M_{2})=\emptyset$. However, on the I-brane there are still anomalies due to the fact that we used only a local form for $\tau_{M}$. 

Such anomalies are described in [12] and the inflow method is used in order to cancel them. 

Consider the case when two D-branes intersect and therefore one obtains massless fermions from the open string sectors with two ends on the two D-branes. Given $N_{1}$ D-branes wrapping around $M_{1}$ and $N_{2}$ D-branes wrapped around $M_{2}$ and the sector of the string starting on $M_{1}$ and ending on $M_{2}$ the difference in the boundary conditions on the two ends of the string modifies its zero point energy and shifts the modes of its worldsheet operators. As a result, the massless fermions are a section of the chiral spinor bundle lifted from 
\begin{equation}
T(M_{1})\cap T(M_{2})\oplus N(M_{1})\cap N(M_{2})
\end{equation}
and in the end the bundle is tensored with the $(N_{1},\bar{N}_{2})$ vector bundle due to their Chan-Paton quantum numbers. The anomaly can be written as 
\begin{widetext}
\begin{equation}
I_{I-brane}=2\pi \int_{M_{12}}(ch(F_{1})\wedge ch(-F_{2})\wedge \frac{\hat{A}[T(M_{1})\cap T(M_{2})]}{\hat{A}[N(M_{1})\cap N(M_{2})]}\wedge e[N(M_{1})\cap N(M_{2})])^{(1)}
\end{equation}
\end{widetext}
by means of brane currents we have for the case of intersections
\begin{equation}
I_{I-brane}=\pm 2\pi \int \tau_{M_{12}}\wedge(e[N(M_{1})\cap N(M_{2})]\wedge ch(F_{1})\wedge ch(-F_{2})\wedge \frac{\hat{A}[T(M_{1}T(M_{2})}{\hat{A}[N(M_{1})\cap N(M_{2})])}^{(1)}
\end{equation}
where use has been made of the fact that $e(\emptyset)=1$. It can be checked that the previous equation can be factorised. If we denote 
\begin{equation}
Y_{i}=ch(F_{i})\wedge \sqrt{\frac{\hat{A}[T(M_{i})]}{\hat{A}[N(M_{i})]}}
\end{equation}
together with 
\begin{equation}
\tilde{Y}_{j}=-(-1)^{\frac{dim(M_{j})-q}{2}}ch(-F_{j}) \sqrt{\frac{\hat{A}[T(M_{j})]}{\hat{A}[N(M_{j})]}}
\end{equation}
with this definition, the anomaly can be cancelled by the inflow.

This yields 
\begin{equation}
I_{I-brane}=-\pi \int \tau_{M_{1}}\wedge \tau_{M_{2}}\wedge (((-1)^{\frac{dim(M_{2})-q}{2}}ch(F_{1})\wedge ch(-F_{2})+\{1\leftrightarrow 2\} )\wedge \frac{\hat{A}[T(M_{1})\cap T(M_{2})]]}{\hat{A}[N(M_{1})\cap N(M_{2})]})^{(1)}
\end{equation}
It is clear that the two terms in the integrand above sum up rather than cancelling each other, leading to the anomaly 
\begin{equation}
I_{I-brane}=-(-1)^{\frac{dim(M_{2})-q}{2}} 2\pi \int \tau_{M_{1}}\wedge \tau_{M_{2}}\wedge (ch(F_{1})\wedge ch(-F_{2})\wedge \frac{\hat{A}[T(M_{1})\cap T(M_{2})]]}{\hat{A}[N(M_{1})\cap N(M_{2})]})^{(1)}
\end{equation}
While factorizability is important for inflow anomaly cancellation, when the relevant normal bundle is nontrivial, the Euler class in the integrand of $I_{I-brane}$ makes it non-factorizable. The cause for this is the fact that while $\tau_{M}$ is a physical observable, it is not always globally defined over $M$ [12]. 
\begin{equation}
\tau_{M}=\delta(x^{1})dx^{1}\wedge ... \wedge \delta(x^{dim N(M)})dx^{dim N(M)}
\end{equation}
makes sense only within each coordinate patch. Between patches the transversal coordinates are defined only up to the transition functions of the normal bundle. We therefore need to add new terms which vanish when $N(M)$ is trivial but allow us to define $\tau_{M}$ globally when $N(M)$ is not trivial. These new terms will carry the topological information about $N(M)$ and, according to the defining equation of $\tau_{M}$ they must have components with indices tangential to $M$. 
This calculation has been covered in [12] and I repeat it here for completeness. When pulling $\tau(M)$ back to $M$ only certain pieces of the correction remain. The result will be the Euler class $e[N(M)]$ of $N(M)$. It may be noticed that $\tau_{M}$ is determined by $N(M)$ because it is defined as the limit of non-singular differential forms with shrinking compact supports in the neighbourhood of $M$, approximated by the neighbourhood of the zero section of $N(M)$. For any oriented real orientable vector bundle $E$, we can define $\tau_{M}$ by taking $M$ to be the zero section $E$. We may define $\Phi[E]=\tau_{M}$, for any vector bundle $E\xrightarrow{\pi} M$. Otherwise stated 
\begin{equation}
\tau_{M}\wedge \tau_{M}=\tau_{M}\wedge \Phi[N(M)]=\tau_{M}\wedge [e[N(M)]]
\end{equation}
where $[e]$ denotes a representative of the cohomology class of $e$. Given also that $\Phi(A\oplus B)=\Phi(A)\wedge \Phi(B)$, we have for the $I$-brane worldvolume $M_{12}=M_{1}\cap M_{2}$ that 
\begin{widetext}
\begin{equation}
\begin{array}{r}
\tau_{M_{1}}\wedge \tau_{M_{2}}=\Phi[T(M_{1})\cap N(M_{2})\oplus N(M_{1})\cap N(M_{2})]\wedge \Phi[N(M_{1})\cap T(M_{2})\oplus N(M_{1})\cap N(M_{2})]=\\
=\Phi[T(M_{1})\cap N(M_{2})\oplus N(M_{1})\cap T(M_{2})\oplus N(M_{1})\cap N(M_{2})]\wedge \Phi[N(M_{1})\cap N(M_{2})]=\\
=\tau_{M_{12}}\wedge e[[N(M_{1})\cap N(M_{2})]]
\end{array}
\end{equation}
\end{widetext}
This is finally the correct replacement for the naive equation 

\begin{equation}
\tau_{M_{1}}\wedge \tau_{M_{2}}=\left\{
    \begin{array}{ll}
        \tau_{M_{12}} & \mbox{if } N(M_{1})\cap N(M_{2})=\emptyset \\
        0 & \mbox{otherwise}
    \end{array}
\right.
\end{equation}

Therefore, there exists an additional gauge-type symmetry in quantum field theories, in particular such quantum field theories that rely on a cohomology description. Aside the usual gauge degrees of freedom, we have yet another freedom of changing the theory, this time by altering the coefficient structure while following the exact sequence given by the universal coefficient theorem. 
This "gauge" symmetry is extremely powerful, as various relations can emerge between apparently different theories that are however connected by means of universal coefficient theorems. This type of "higher" symmetry has been discussed previously in [19]. In this article I show that such a higher symmetry can be used in order to eliminate certain anomalies emerging in the context of intersecting branes. 


The main idea of this article is that the same information can be encoded by coefficient structures in (co)homology. 
A different way of saying this is to observe that homology in general represents a partition of a topological space. The homology groups partition the topological space under study into homology classes (finally equivalence classes) that are designed to retrieve certain geometrical data. This in turn allows us to differentiate between topological spaces based on the criterium that they share the same type of geometrical data as that retrieved by the homology. The dualization of this construct leads to cohomology, which, while it looks similar, it often encodes more or different types of information than homology. 

It is attractive to see the coefficient structure as a form of gauge, and to see a transformation from one coefficient structure to another as a gauge transformation of a special kind. In fact, I am using this interpretation in this article. But the coefficient structure does more than this. As said, homology offers a sort of partitioning of the topological space, in such a way that a specific type of geometric data can be retrieved, while making other types of data blurred or undetermined. This reminds us more of the compatibility of observables in quantum mechanics, but then, I showed relatively recently in ref. [20], [21] that the two, namely gauge freedom and quantum mechanics may have some aspects in common.

While the topological space to be probed remains the same, there is an intrinsic limitation to the type of geometrical information that can be retrieved from it, and that limitation is given by the coefficient structure in (co)homology. This structure basically encodes the type of "partition" of our topological space, which makes some of the geometrical or topological properties manifest, while obscuring others. 
The "partitioning" however needs to be understood in a more general sense, as changing the coefficients of (co)homology amounts to changing the point structure used to describe the space. Basically, having coefficients different from $Z$ means modifying the Eilenberg-Steenrod axioms. In the ordinary (co)homology case, the dimension axiom states that for $P$ being a one-point space, the homology $H_{n}(P)=0$ for all $n\neq 0$ and if $P$ is a one-point space $H_{0}(P)$ is the coefficient group. This coefficient group is for ordinary (co)homology the integers. This would imply a partitioning of the topological space we wish to describe by means of geometric points with the usual structure of a point. If this axiom is abandoned however, the partitioning will be done by means of a different object, maybe a string, or a brane, in any case, a structure that would be sensitive to geometric and topological features of the space to which our ordinary (co)homology would be not. It is possible to imagine that a symmetry relating different coefficient structures (by means, for example, of the universal coefficient theorem) would reveal at the level of the analysed space, higher symmetries that could not be otherwise detected. 
These same coefficient structures have the ability to carry an anti-anomaly inducing the cancellation of the original anomaly by altering the pairing between cohomology and homology, that is, altering the integral over the whole space $X$ involving $\tau_{M}$. In a sense the "inflow" can be seen as originating in the prescription of altering the pairing 
\begin{equation}
<,>:H^{q}(X;A_{2})\times H_{q}(X;A_{1})\rightarrow R
\end{equation}
such that its adjoint becomes part of an exact sequence. This leads to a global definition of the pairing and particularly in the ability to define $\tau_{M}$ such that it is valid over the whole space. As stated before, the conditions for the sequence to be exact lie in $Ext_{R}(H_{q-1}(X;A_{2}),A_{1})$. Let me analyse this a bit further. To understand the concept of extension one has to imagine exact sequences as self-sustained mathematical entities. Indeed, one may act on an exact sequence $0\rightarrow A\rightarrow B\rightarrow C\rightarrow 0$ with a covariant functor $F$. The functor may be itself right (resp. left) exact if, after applying it to the exact sequence one obtains a sequence $F(A)\rightarrow F(B)\rightarrow F(C)\rightarrow 0$ (respectively $0\rightarrow F(A)\rightarrow F(B)\rightarrow F(C)$) which is exact. When the functor with which we act on our exact sequence is $Hom_{R}(*,M)$ where $R$ is a Ring and in general $M$ is a module then the sequence 
\begin{equation}
0\rightarrow Hom_{R}(C,\mathcal{A})\rightarrow Hom_{R}(B,\mathcal{A})\rightarrow Hom(A,\mathcal{A})
\end{equation}
is not exact. In order to make it exact what we need to add are the extensions, which, by definition, transform the resulting sequence into 
\begin{equation}
\begin{array}{c}
0\rightarrow Hom_{R}(C,\mathcal{A})\rightarrow Hom_{R}(B,\mathcal{A})\rightarrow Hom(A,\mathcal{A})\rightarrow \\
\\
\rightarrow Ext_{R}^{1}(C,\mathcal{A})\rightarrow ... \rightarrow Ext_{R}^{q}(B,\mathcal{A})\rightarrow Ext_{R}^{q}(A,\mathcal{A})\rightarrow Ext_{R}^{q+1}(C,\mathcal{A})\rightarrow ... \\
\end{array}
\end{equation}
which is exact. The extension satisfies the property that $Ext_{R}^{0}(\mathcal{A}_{1},\mathcal{A}_{2})=Hom_{R}(\mathcal{A}_{1},\mathcal{A}_{2})$ and $Ext^{n}_{R}(F,\mathcal{A})=0$ if $F$ is a free module and $n>0$. Therefore, $Ext_{R}^{n}(*,\mathcal{A})=0$ is called the $n$-th derived functor of the functor $Hom_{R}(*,\mathcal{A})$. Therefore, it is intuitive to see that $Ext$ represents what needs to be added to the resulting sequence after the application of a functor to it, such that the resulting sequence becomes exact. Now, of course, cohomology theory measures (in its most basic form) in how far a complex fails to be exact. Therefore, in our case, the case of the universal coefficient theorem, we obtain an exact sequence in (co)homology after changing the coefficients. The sequence we obtain is certainly cohomologically trivial (being exact) and this is obtained due to the fact that we added the anti-anomaly as a structure given to the mathematical points of our theory. The addition of $Ext$ therefore makes our $\tau_{M}$ well defined over the whole space without having to look at the properties of higher dimensional spaces. Obviously, as $\tau_{M}$ is an element of a cohomology class, we discuss exact sequences in cohomology and apply the $Hom$ functor to those, while changing their coefficients. This basically demands for the universal coefficient theorem. 


For cohomology, the universal coefficient theorem states that the following sequence is exact:
\begin{equation}
0\rightarrow Ext(H_{p-1}(X;\mathcal{A}_{1}),\mathcal{A}_{2})\rightarrow H^{p}(X;\mathcal{A}_{2})\xrightarrow{h} Hom(H_{p}(X;\mathcal{A}_{1}),\mathcal{A}_{2})\rightarrow 0
\end{equation}
Given two coefficient structures $\mathcal{A}_{1}$ and $\mathcal{A}_{2}$ this short exact sequence connects cohomology with coefficients in one structure to homology with coefficients in the other. If the map between the two were an isomorphism, the $Ext$ group would be trivial. This is often not the case, particularly when one wishes to connect cohomology theories induced by coefficient structures that have different resolutions with respect to certain topological features. This is precisely the situation I discussed above, considering that the RR charges will not be visible in the perturbative domain of the string theory. Indeed, it is not possible to use cohomology to detect topological features without making a choice of a coefficient structure. Such  choice expands or reduces the structure associated to the fundamental object (the point in this case) and at the same time makes certain topological features more or less visible. Indeed, this is valid for K-theory which is, axiomatically speaking, also a generalised cohomology theory. Another aspect related to changing the coefficient structure in cohomology is related to the relativity of symmetry, a property I proved in ref. [11], (Theorem 3, pag. 4). It is intriguing that a special case of this result has been obtained [3] by a different and more particular method. Indeed, [3] notices that M-theory with $N>0$ supersymmetry can be seen in terms of type IIA perturbative string theory as having $N=0$ as long as $RR$ charge is not detectable. The tension point of [3] is between the so called "revolutionary" and "counter-revolutionary" viewpoints. While the "revolutionary" viewpoint will claim that the 11-dimensional M-theory is fundamental, and any vacuum of the 11-dimensional supergravity is acceptable, independent of the results of a low energy type IIA theory, while the "counter-revolutionary" viewpoint will claim that the only acceptable vacua are those of type IIA supergravity, while M-theory is only a strong coupling limit of the string theory. The tension between these two viewpoints can now be alleviated by understanding that the two situations arise due to the choice of two different coefficient structures in cohomology, one capable of revealing the topological features associated to the RR-charge and the other incapable of doing so. The two choices are however perfectly legitimate from the perspective of the universal coefficient theorem which allows us to move from one description to the other.  In reference [13] D-brane Chern-Simons actions were derived. These imply the presence of topological defects on the D-branes. Such defects carry their own RR charges which are determined by their own topological (instanton) numbers [14]. Anomaly cancelation arguments by the inflow mechanism [14] require the modification of the Chern-Simons actions. Of course the effect will be to change the induced RR charges on the D-brane wrapped around a cycle of a non-trivial compactification manifold. In terms of cohomology with non-trivial coefficients this can be interpreted as adding the additional structure to points. Of course, in the most general case, adding or removing additional topological structure (additional topological defects) can be done for arbitrary extended objects like D-branes but also, as is the case here, for Dirichlet 0-branes. 
The author of ref. [3] concludes that the possibility of noticing supersymmetry in the non-perturbative theory and "overlooking" it in the perturbative approach could be a method to "have your supersymmetry and eat it too". However, the final conclusion is that nature should intrinsically be non-perturbative and therefore what one must measure is the non-perturbative physics including supersymmetry and not merely the perturbative region of some string theory. Having supersymmetry and eat it too, according to ref. [3] can only occur if "for some unknown reason, the experimentalist's apparatus is so primitive as to be unable to detect Ramond-Ramond charge, in which case he or she would conclude that the world has no unbroken supersymmetry". Analysing this situation from the perspective of cohomology with non-trivial coefficients and the universal coefficient theorem, it results that, in order to detect certain topological features one has to make a choice of coefficients that may reveal or hide topological features associated to the Ramond-Ramond charge, leading to visibility resp. invisibility of the latter. Therefore, the experimentalist might not be able to detect certain topological features unless he or she is willing to create an experiment in which some other features will remain undetectable. This way of thinking is well known from basic quantum mechanics where incompatible observables cannot have a common eigenbasis. To extend this to the level of detectability of topological features one requires the universal coefficient theorem. 
Formally, one has to consider the changes leading to $\tau_{M_{1}}\wedge \tau_{M_{2}}=\tau_{M_{12}}\wedge e[[N(M_{1})\cap N(M_{2})]]$ from a purely cohomological perspective. The change in I-brane current $\tau_{M_{1}}\wedge \tau_{M_{2}}$ of an I-brane measuring intersections of the two branes is naively trivial when $N(M_{1})\cap N(M_{2})\neq \emptyset$. As an observable $\tau_{M}$ must be globally defined over $M$. When $N(M)$ is non-trivial one needs to change $\tau_{M}$ topologically such that it becomes globally well defined. Indeed, this can be done by encoding the full topological information within the coefficients in cohomology. I showed on several occasions that the coefficients in cohomology can play the role of global anti-anomalies or of structures that may turn anomalous theories into well defined ones. 
\section{Examples and discussion}
The inflow anomaly cancellation usually refers to the mechanism through which an anomalous contribution in one dimension, usually on a boundary, is cancelled by a contribution from the "bulk" manifold. This anomaly is not only important in string theory and in particular in the case of intersecting D-branes but also when we consider for example Dirac operators on various types of manifolds. In fact, this subject has been studied in the context of topological phases of matter as well as in various analyses of quantum field theory. For recent discussions see [22]-[28] and for a historical approach to the matter the reader can look into [29]. 

In particular it is useful to distinguish anomalies that can be eliminated by inflow as being perturbative and non-perturbative. It is well known that perturbative anomalies in a spacetime of dimension $d$ are related to a Chern-Simons function in dimension $d+1$. A condensed matter example for this type of inflow anomaly cancellation is the integer quantum Hall effect. However, in the description of fermions on manifolds with or without boundary, we encounter not only perturbative anomalies but also non-perturbative ones. In order to describe systems with such non-perturbative anomalies, the Chern-Simons functional term is replaced by the $\eta$-invariant of the Atiyah-Patodi-Singer theorem.
Another situation is the one related to anomalous bosonic systems. Condensed matter systems often present a so called topologically protected symmetry (SPT) phase in which we have a gapped phase with a global symmetry group $G$, and with the property that the phase is topologically non-trivial but would become trivial if the symmetry was explicitly broken. We can therefore re-interpret the anomaly inflow in this context to better understand the SPT phases, provided that the role of the $\eta$ invariant of the non-perturbative fermion anomaly is replaced with group cohomology. In all cases, as has been noted in ref. [23] the anomaly in dimension $d$ is related to an invertible topological field theory in dimension $d+1$. 
The general idea is that a gauge anomaly emerging in a lower dimension (for example on a boundary) is being cancelled by a similar inverse anomaly emerging in a higher dimension (say, in the associated bulk space of that boundary). The inflow anomaly that cancels the lower dimensional anomaly usually appears due to topological features of the higher dimensional manifold (unless we have a perturbative anomaly which can be explained however also by means of Chern-Simons counter-contributions). 
What I showed in this article was that we can identify the relative freedom of choosing a coefficient structure in cohomology (to be compensated with $Tor$ and $Ext$ groups in the exact sequence of the universal coefficient theorem) with an additional gauge symmetry, one previously unknown, that, when implemented as such, results in additional terms that would annihilate the usual anomalies in the lower dimensional manifold. 
These terms will appear in the construction of the inner products (our bra-ket combinations we use often in quantum mechanics) as modifications due to the additional segments of the exact sequence of the universal coefficient theorem. In fact, as quantum states are perfect indicators of the non-local and topological features of the manifolds on which they are defined, it makes perfect sense to encode topological information at the level of bra and ket states. The associated pairing involving Dirac operators therefore will lead us to the action functionals and partition functions for the fermion system we wish to describe. However, if the background context involves topological features that are non-trivial, in particular, such feature that can be described by cohomology with various coefficients and we have to impose a coefficient structure invariance by using the universal coefficient theorem (and hence account for the $Tor$ and $Ext$ terms) then our pairings will get a different form, one that will modify the action functional and the partition function accordingly. These modifications are found to be of the type required to cancel the inflow anomaly terms. There are several types of pairings between homology groups and between homology and cohomology, and as quantum states are naturally sensible to such topological features, it makes sense to encode the (co)homological information in the form of bra and ket representations. In principle the homology groups $H_{k}(C)$ relate to the shape of the manifold. The cohomology groups $H^{k}(C)$ relate to the differential forms defined over the manifold. Hence, if there is a manifold, say, $M$, characterised by a sequence of homology groups then one can define the integral 
\begin{equation}
\int_{M}\omega
\end{equation}
characterised by the differential form $\omega$ and by the manifold $M$. Integration can be seen as the pairing 
\begin{equation}
H_{k}(M, \mathbb{R})\times H^{k}(M,\mathbb{R})\rightarrow \mathbb{R}
\end{equation}
such that we have
\begin{equation}
([M],[\omega])\rightarrow \int_{M}\omega
\end{equation}
where this pairing is constructed with real coefficients and this coefficient structure characterises also the measure of integration and implicitly the differential form $\omega$. Here, $[M]$ represents a class in homology and $[\omega]$ represents a class in cohomology. The pairing above however is an isomorphism (one-to-one relation) only when this particular choice of coefficients is made. For other coefficients this pairing may fail to be an isomorphism. The correction is encoded in a term controlled by the $Ext$ and $Tor$ groups
\begin{equation}
H_{k}(M,\mathbb{G})\times H^{k}(M,\mathbb{G})\rightarrow \mathbb{G}
\end{equation}
where the map becomes 
\begin{equation}
([M],[\omega])\rightarrow \int_{\{M\}}\omega \oplus \int_{Ext(H_{n-1}(C),\mathbb{G})}\eta
\end{equation}
Here the first integral is over the manifold $\{M\}$, visible when the coefficient structure $\mathbb{G}$ is used and the correction appears as an integral of another differential form over the extension group constructed from the homology with general integer coefficients over a lower dimension. Here, I simply used the universal coefficient theorem in cohomology. The non-trivial topology however is not visible from the lower dimension, and hence the simplification, but this simplification is compensated by the extra terms appearing in the action (the anomaly terms emerging from the requirement of coefficient choice invariance). This shows that properties defined on some more complex topological objects may be acceptably described on simpler topological objects if controlled changes in the groups used to describe them are being employed. We see therefore that in the process of calculation of pairings, the topological structure encoded in the ket and bra vectors can be moved into the description of the pairings, and that imposing invariance with respect to changes in coefficients implies the introduction of additional terms in the process of taking the inner product. In fact, in the description of the spectra of Dirac operators, the encoding of topological features in ket and bra vectors appears quite often in ref. [23] and [24], and that for a very good reason. My argument here is that the construction of partition functions in the form of pairings between fermion states as expressed in [23] and [24] is dependent on specific choices of coefficients in cohomology, in particular because the topological information encoded in such states is in fact (co)homological in nature. Therefore the emergence of a non-perturbative anti-anomaly encoded by the $\eta$-invariant can be related to a modification of the pairing used to construct the partition function, originating from the application of the universal coefficient theorem between different coefficient structures in cohomology. Requiring invariance with respect to such choices of cohomology results in correction factors appearing in the form of the $\eta$-invariant. 
Let us therefore start with a massive Dirac fermion $\Psi$ on a $D$-dimensional manifold $Y$ with charges from $U(1)$. We also presume the existence of a conjugate field $\bar{\Psi}$ that carries the opposite $U(1)$ charges. The general action for such a fermion can be written as 
\begin{equation}
I=-\int_{Y}d^{D}x\sqrt{g}\bar{\Psi}(\slashed{D}_{Y}+m)\Psi
\end{equation}
Here we have a Dirac operator defined on the higher dimensional manifold $Y$, with the form 
\begin{equation}
\slashed{D}_{Y}=\sum_{\mu = 1}^{D}\gamma^{\mu}D_{\mu}
\end{equation}
with $\{\gamma_{\mu},\gamma_{\nu}\}=2 g_{\mu\nu}$. If we assume the manifold $Y$ has no boundary, then the Dirac operator $D_{Y}=i\slashed{D}_{Y}$ is self-adjoint. 

If, however, the manifold $Y$ has a boundary, $W=\partial Y$ then we can represent the metric of $Y$ as 
\begin{equation}
ds_{Y}^{2}=d\tau^{2}+ds_{W}^{2}
\end{equation}
where $ds_{W}^{2}$ is the metric on the boundary with the time coordinate $\tau$ parametrising the normal direction. We may adopt the normalisation of [23] that $\tau$ is zero on the boundary and negative away from the boundary inside $Y$. 
Our massive fermion $\Psi$ will have to obey a boundary condition in the case of a manifold with boundary. This boundary condition will have to satisfy several properties according to ref. [23]. In particular the boundary condition is required to be elliptic but not self-adjoint. Ellipticity of the boundary condition ensures that the Dirac operator with such a boundary condition has the properties required to make the Euclidean theory well defined [23]. In particular a boundary condition on the Dirac equation at $\tau=0$ is elliptic if, after dropping from the equation all lower order terms like mass terms and couplings to background fields, or after taking the momentum along the boundary to be very large, the equation and its adjoint have no solutions that satisfy both our elliptic boundary condition and vanish for $\tau\rightarrow -\infty$. 
The boundary condition we therefore can impose will be 
\begin{equation}
L:(1-\gamma^{\tau})\Psi_{\tau=0}=0
\end{equation}
where $\gamma^{\tau}$ is the gamma matrix in the $\tau$ direction. 
We can see that this boundary condition is elliptical but it is not self-adjoint. In fact, this is a desirable property because any attempt to create a self-adjoint boundary condition will fail due to the surface term at $\tau=0$. Moreover a self-adjoint boundary condition would not lead to a chiral potentially anomalous system which we want to study [23]. 
However, we still will need, if we are to describe a physically meaningful system on the boundary $W$, that the dynamics of the boundary modes is described by a self adjoint Hamiltonian. If we continue $W$ to the Lorentz signature and keep $x^{0}$ as the time coordinate, then the Hamiltonian that propagates a state in the $x^{0}$ direction must still be self-adjoint, and the boundary condition $L$ satisfies this requirement. 
Therefore we can re-write the action functional near the boundary as 
\begin{equation}
I=-\int_{Y}d^{D}x\sqrt{g}\bar{\Psi}\gamma^{\tau}(\frac{\partial}{\partial \tau}+D_{W}+\gamma^{\tau}m)\Psi
\end{equation}
where
\begin{equation}
D_{W}=\sum_{\mu\neq\tau}\gamma^{\tau}\gamma^{\mu}D_{\mu}
\end{equation}
is a self adjoint Dirac operator on $W$. We call the operator $\gamma^{\tau}$ a chirality operator. Following the rest of the constructions from ref. [23] one can easily observe that the Dirac equation has a mode localised near the boundary for $m<0$ and therefore we can write 
\begin{equation}
\begin{array}{ccc}
\Psi=\chi exp(|m|\tau), & (1-\gamma^{\tau})\chi=0, & D_{W}\chi =0\\
\end{array}
\end{equation}
with $\chi$ a fermion field on the even dimensional boundary $W$. The equation satisfied by this fermion is $D_{W}\chi=0$ and hence the fermion mode propagating along $W$, namely $\chi$, is massless. It also obeys the equation $\gamma^{\tau}\chi=\chi$ which makes this fermion chiral. After regularisation we can easily quantise this fermion. 
If we want to calculate the partition function of the massive fermion $\Psi$ on the manifold $Y$ with the above boundary condition $L$ we may imagine the coordinate $\tau$ as a Euclidean time coordinate and the boundary $W=\partial Y$ as a time slice. The path integral over $Y$ provides us with a physical vector state which we call $\ket{Y}$. This would be a vector state in the Hilbert space $\mathcal{H}_{W}$ of $W$. The boundary condition is also a state vector in $\mathcal{H}_{W}$ namely $\ket{L}$.
The partition function is therefore written as a pairing between these two states in the form 
\begin{equation}
Z(Y,L)=\bra{L}\ket{Y}
\end{equation}
In this construction the respective states encode also the topological features of both the bulk space and the boundary, as well as the topological structure of the boundary condition. In fact, as shown in [25], the boundary condition itself is non-local. The pairing introduced above in the form of an inner product encodes also topological information in such a way that the end result preserves a series of symmetries, in particular gauge symmetries. However, at this point nothing has been said about the symmetry resulting from the changes of coefficients in cohomology, and it is precisely there where topological corrections may appear, in the form of $Tor$ and $Ext$ contributions. 
If we introduce a large mass gap in the Hilbert space $\mathcal{H}_{W}$ as implemented by [23], by considering the mass $|m|$ very large, the path integral on the cylindrical region $(-\epsilon, 0]\times W$ gives an Euclidean time evolution $e^{-\epsilon H}$ where $H$ is our Hamiltonian. With a very large mass, the factor $e^{-\epsilon H}$ plays the role of a projector on the ground state $\ket{\Omega}$:
\begin{equation}
e^{-\epsilon H}\sim \bra{\Omega}\ket{\Omega}
\end{equation}
and in this limit $\ket{Y}\sim \ket{\Omega}$ with the vacuum state normalised $\bra{\Omega}\ket{\Omega}$. 
This will allow us to write the partition function as 
\begin{equation}
Z(Y,L)=\bra{L}\ket{\Omega}\bra{\Omega}\ket{Y}
\end{equation}
This allows us to split the bulk part from the boundary part (encoded by the boundary condition). 
The ground state $\ket{\Omega}$ carries a phase ambiguity. The space of ground states $\mathcal{L}_{W}$ is a one dimensional subspace of the Hilbert space $\mathcal{H}_{W}$. Let the background fields be characterised by a parameter space $\mathcal{W}$. If we transport the vacuum state in a loop on the parameter space of the background, there is a non-trivial holonomy which translates into a Berry phase for our quantum state. Each point on the parametric space $\mathcal{W}$ combines with the one dimensional space $\mathcal{L}_{W}$ resulting in a complex line bundle over $\mathcal{W}$ with non-trivial holonomies determined by the Berry phases. As we can easily see, the numbers $\bra{\Omega}\ket{Y}$ and $\bra{L}\ket{\Omega}$ which do not have a well defined phase can be replaced with the following objects
\begin{equation}
\begin{array}{cc}
\ket{\Omega}\bra{\Omega}\ket{Y}\in \mathcal{L}_{W}, & \bra{L}\ket{\Omega}\bra{\Omega}\in \mathcal{L}^{-1}_{W}
\end{array}
\end{equation}
which are well defined as they do not depend on the phase of the state $\ket{\Omega}$. 
However, this replaces the usual scalar inner product with a vector and it has a dynamics inside a fibre bundle. This amounts to the fact that we have to understand the partition function of the chiral fermion $\chi$ on the boundary $W$ as a section of a line bundle $\mathcal{L}_{W}^{-1}$ also known as the determinant line bundle and not a simple complex number. The factor $\bra{\Omega}\ket{Y}$ is an exponentiated $\eta$-invariant that can be seen as an element of $\mathcal{L}_{W}$. 
If the Dirac operator on $W$ has no zero modes, we can still avoid the line bundle description if we replace the vacuum state $\ket{\Omega}$ with one that has no phase ambiguity. We can in fact use the so called Atiyah-Patodi-Singer boundary condition noted in [23] "APS" where, with such a condition, the path integral on $Y$ corresponds to a state vector $\ket{APS}\in\mathcal{H}_{W}$. With $\ket{Y}$ a multiple of $\ket{\Omega}$ we obtain the partition function as 
\begin{equation}
Z(Y,L)=\frac{\bra{L}\ket{\Omega}\bra{\Omega}\ket{APS}}{|\bra{APS}\ket{\Omega}|^{2}}\cdot \bra{APS}\ket{Y}
\end{equation}
where the inner product $\bra{APS}\ket{Y}$ is the partition function of the massive fermion $\Psi$ on $Y$ with APS boundary condition. In the limit of $Y$ and $W$ large compared to the massive fermion Compton wavelength, we have
\begin{equation}
\bra{APS}\ket{Y}=exp(-i \pi \eta_{D})
\end{equation}
where $\eta_{D}$ is the invariant defined by Atiyah-Patodi-Singer. 
The remaining factor is simply the determinant of the Dirac operator on the boundary $|Det(D_{W}^{+})|$ for the massless fermion $\chi$ (with positive chirality here). 
Therefore
\begin{equation}
Z(Y,L)=|Det(D_{W}^{+})|\cdot exp(-i \pi \eta_{D})
\end{equation}
This is the standard approach analysed in particular in [23-28] and is the basis of the inflow anomaly calculations for situations with non-trivial topology. However, to a similar result one can arrive if one analyses what happens when, in the context of the line bundles of the type $\mathcal{L}_{W}$, we change the coefficients in cohomology that are supposed to detect such topological structure. What we have here is an initially cyclical manifold where acyclicity emerges due to the line bundle structure encoded via the phase ambiguity. We can eliminate it by either setting up a non-local boundary condition like the APS boundary condition, and hence maintaining the initial scalar product, or we can keep the line bundles, work with the initial boundary conditions and the state $\ket{\Omega}$ but then correcting for the acyclicity emerging from this approach. By acyclicity we mean basically the fact that in general, cyclical coefficient structures used to detect topological features make the topological "detector" (here, the cohomology) insensitive to cyclical structures. Therefore, using a cyclical coefficient structure makes our manifold (where a Berry phase exists) equivalent with a situation in which no non-trivial Berry Phase holonomy exists, and no different boundary conditions are required. However, this transition doesn't occur at no cost. In fact, the cost is precisely a Torsion type correction of the inner product, which is by our construction, cyclical, namely 
\begin{equation}
exp(-i \pi \eta_{D})
\end{equation}
The way in which cyclic coefficients can hide information in (co)homology, which however can be detected when changing coefficients, I will describe by referring to my own article ref. [26]. Following that, let me consider a complex bundle, which in particular can represent also quantum states. The connectivity of such a space is determined by means of (co)homology with, say, complex coefficients. When the algebraic structure of the coefficients in cohomology is modified, the accessible information about the connectivity of our space may change. A specific non-trivial choice of coefficients can therefore lead to a non-trivial superposition of disconnected topological spaces which may result in a connected topological space. Let me therefore consider two circular spaces $S^{1}$ as explained in [26] and show that by a particular change in coefficients, the two circular spaces representing together a disconnected space will become homeomorphic to a single space and hence connected (albeit not simply connected). Then, the resulting non-simply connected space will be mapped by another change of coefficients into a simply connected space homeomorphic to a single point. The translation of this effect in the current problem is that the topologically non-trivial effects encoded by the line bundle can be eliminated by modifications in the coefficients of cohomology, while demanding the invariance of the physics with respect to such changes amounts to the emergence of Torsion terms in the universal coefficient theorem which modify our inner products by means of Tor integrals presented above. For this to occur, the particular choice of coefficients must contain a certain twisted cyclicality. This would correspond to our Berry phase in the present problem. For simplicity I will discuss the process in terms of integer and twisted cyclical integer coefficients. This will also suit very well in the construction of the APS boundary condition and the APS invariant $\eta$. Let us therefore start with a circle space $S^{1}$ and an abelian group $A$. Then let us define 
\begin{equation}
\rho:\pi_{1}S^{1}\rightarrow Aut(A)
\end{equation}
a representation of the fundamental group of the circle into the abelian group $A$. Then, the homology of the circle with coefficients in the group $A$ twisted by the map $\rho$ is $H_{k}(S^{1},A_{\rho})$. If one considers as I did in [26] a simple example of $A=\mathbb{Z}_{3}$ and the map $\rho:\mathbb{Z}\rightarrow Aut(\mathbb{Z}_{3})$ as being 
\begin{equation}
\rho= \left \{
\begin{aligned}
 0\rightarrow 0 \\
 1\rightarrow 2\\
 2\rightarrow 1\\
 3\rightarrow 0\\
 4\rightarrow 2\\
 ...\\
\end{aligned} \right.
\end{equation}
The cellular chain complex associated to the homological representation of the circle is then
\begin{equation}
0\rightarrow \mathbb{Z}[t,t^{-1}]\xrightarrow{\delta}\mathbb{Z}[t,t^{-1}]\rightarrow 0
\end{equation}
where $\delta$ is the boundary map which by definition represents the multiplication with $(t-1)$. This means that $t$ and $t^{-1}$ define the required ring structure for the circular space. Therefore we obtain the isomorphism 
\begin{equation}
\mathbb{Z}[\pi_{1}S^{1}]\cong \mathbb{Z}[t,t^{-1}]\cong \mathbb{Z}[\mathbb{Z}]
\end{equation}
which simplifies the calculation. 
We now tensor with $\mathbb{Z}_{3}$ to obtain the homology with the desired coefficients over $\mathbb{Z}[t,t^{-1}]$. Then we obtain
\begin{equation}
\mathbb{Z}_{3}\xrightarrow{\cong}\mathbb{Z}[t,t^{-1}]\otimes_{\mathbb{Z}[t,t^{-1}]}\mathbb{Z}_{3}\xrightarrow{\delta\otimes Id}\mathbb{Z}[t,t^{-1}]\otimes_{\mathbb{Z}[t,t^{-1}]}\mathbb{Z}_{3}\xrightarrow{\cong}\mathbb{Z}_{3}
\end{equation}
The first map is $a\rightarrow 1\otimes a$ and the last map is $1\otimes a\rightarrow a$. We reduce to $1\otimes a$ before we apply the last map, the result being 
\begin{equation}
a\rightarrow 1\otimes a\rightarrow (t-1)\otimes a = 1\otimes (t a-a)\rightarrow t a-a
\end{equation}
The boundary map obtained after taking the tensor product with $\mathbb{Z}_{3}$ is 
\begin{equation}
D:\mathbb{Z}_{3}\rightarrow \mathbb{Z}_{3}
\end{equation}
\begin{equation}
\begin{array}{c}
D(0)=0\\
D(1)=t\cdot 1 - 1 = 2-1=1
D(2)=t\cdot 2-2=1-2=2
\end{array}
\end{equation}
(due to the nature of $\mathbb{Z}_{3}$) and hence it is the identity on $\mathbb{Z}_{3}$. Therefore the homology groups of $S^{1}$ with coefficients in $\mathbb{Z}_{3}$ twisted by the non-trivial map $\rho$ are all trivial 
\begin{equation}
H_{0}(S^{1};\mathbb{Z}_{3})_{\rho}\cong H_{1}(S^{1};\mathbb{Z}_{3})_{\rho}\cong...\cong 0
\end{equation}
Therefore a circle can be mapped into a point via a controllable change of coefficients in homology. A similar procedure can merge two disjoint circles into one single circle. In order to do this the coefficient group $A$ will now be $\mathbb{Z}_{2}$ and the twisting will have the form 
\begin{equation}
\rho= \left \{
\begin{aligned}
 0\rightarrow 1 \\
 1\rightarrow 0\\
 2\rightarrow 1\\
 3\rightarrow 0\\
 4\rightarrow 1\\
 ...\\
\end{aligned} \right.
\end{equation}
The analysed space will now be a disjoint union of circles $S^{1}$ namely $X=S^{1}\sqcup S^{1}$. Using the Mayer-Vietoris theorem it results that 
\begin{equation}
H_{q}(S^{1})\cong H_{q}(S^{1})\oplus H_{q}(S^{1})
\end{equation}
Using now the twisted coefficients as described above the homology won't be able to distinguish the two circles and hence we arrive at the single circle case. 
Therefore we can see that the coefficient groups in cohomology determine the topological resolution of the cohomology, having the ability to make certain cycles in the manifold invisible. In a sense, the choice of a coefficient structure is similar to a choice of gauge, but ultimately, if we are to restore gauge invariance we have to construct a partition function that does not depend on various choices of coefficient groups, at least on the topological side. 
The path integral of the boundary fermion is, as mentioned above, the determinant
\begin{equation}
Det(D_{W}^{+})
\end{equation}
The fermion on the boundary has a gauge anomaly which makes it a non-gauge invariant construct, but this anomaly appears in a similar way when integrating the massive fermion on the bulk space $Y$. The two anomalies, one originating from the higher dimensional space $Y$ and its fermion, the other from the boundary space $W$ and its fermion cancel each other out. This is what we call an inflow anomaly cancellation. The higher dimensional anomaly "flows into" the lower dimensional boundary fermion and cancels the anomaly already existing there. 

If all the information we had would be that the coefficient structure switches between cyclical and acyclical, this is all we would have, but in fact, we have to deal with Dirac operators. 
The Hirzebruch signature theorem defined for closed manifolds is a special situation of the more general index theorem for elliptic operators. We can define a special operator acting on a space of differential forms which is known to have the index identified with the signature of the manifold
\begin{equation}
Ind(\mathcal{O}(S))=Sign(M)
\end{equation}
where $S$ is the space of differential forms on the manifold $M$.

We can construct an operator (and equivalently a boundary value problem) called the signature operator which acts on spaces of differential forms. If we construct the index of such an operator we obtain by Hodge theory the signature of the manifold. 
As stated above, in the case of manifolds, we might think to develop a similar approach, namely to consider a manifold X with boundary Y and construct a suitable elliptic boundary value problem for the signature operator whose index will be the signature of X. 
 
However, as seen previously, the definition of boundary conditions encounters topological obstructions which are non-zero for the operator associable to the signature. The observation by Atiyah, Patodi and Singer was that instead of looking for local boundary conditions, it is more suitable to look for global boundary conditions, as presented above. Following ref. [27] we can write the operator near the boundary as
\begin{equation}
\sigma (\frac{\partial}{\partial u}+B)
\end{equation}
where $B=\pm(d*-*d)$ is a self adjoint operator on $Y$. The boundary condition requires that the boundary value $\phi|_{Y}$ lies in the subspace spanned by the eigenfunctions $\phi_{\lambda}$ of $B$ with $\lambda$ negative. Let $P$ be the orthogonal projection onto the space spanned by the eigenfunctions of $B$ with positive or zero eigenvalues $\lambda \geq 0$, then the boundary condition is simply 
\begin{equation}
P(\phi|_{Y})=0
\end{equation}
The operator $P$ is pseudo-differential and has a symbol $p(y,\xi)$ which is an idempotent matrix $m\times m$ of rank $\frac{1}{2}m$ defined on the cotangent sphere bundle of $Y$. 
The symbol of an operator is usually a scalar or matrix function associated to an operator, that has properties that reflect the properties of the original operator. The symbol of the operator usually takes values in an algebraic structure that is simpler than that of the original operators. If the operators act on function spaces, for example on a function of $n$ variables, or, in general functions on $n$-dimensional manifolds, then the symbol can be constructed as a function of $2n$ variables or functions on a $2n$ dimensional manifold. The relation between symbols and operators is at the foundation of quantum mechanics, where the operator itself corresponds to a quantum observable, while the symbol corresponds to a classical observable. As described above the construction of the boundary condition appears in the form of an operator which has a symbol that encounters obstructions to the implementation of the elliptic boundary condition. The global boundary condition can however be implemented by a global operator (the equivalent of a quantum operator) $P(\phi|_{Y})=0$ which results in a proper generalisation of the finite index and its relation to the signature of the manifold. However, this construction implies that the boundary condition has to be global and implemented, as done before, by a quantum state obtained via the application of a quantum operator. The observation of this article is that in fact if such operators can be defined in a specific cohomology, then altering the coefficient structure has the same effect as the one of imposing a global condition. In particular, requiring that the formulation of our theory is invariant with respect to changes of coefficient structures (of a certain type, say cyclical coefficients to normal ones) leads to moving the topological features from the operators to the form of the inner products used in the construction of the partition function. 
Basically, by this invariance we obtain as a result a shift of the information from the topology side of the problem to the elliptic spectrum side, and hence to spectral information contained in the eigenvalues of our operator. 
In this sense, demanding invariance to changes of coefficients is equivalent to the idea that topological information can both be encoded in the properties of operators acting on differential forms and in the spectrum of our elliptic operators. 
This may seem just like another way of looking at the Atiyah-Singer index theorem, but it is much more than that. The principle of coefficient change invariance introduces a much broader type of gauge symmetry in physics problems, one in which problems unsolvable with one coefficient structure become solvable with another. 
An intriguing application that I was thinking at when writing this article was an alternative way of looking at the cryptographic protocols based on elliptic curves. A symmetry linking cohomology theories with different (elliptic curve) coefficients would allow the mapping of cryptographic hard problems to cryptographic simple ones. Moreover, the universality of such a new symmetry implies that such transformations can always be performed. Invariance to choices of coefficients must be a universal property and therefore, for any hard elliptic curve cryptographic problem there must be an equivalent trivial problem "out there". Finding that connection however is not trivial. This however, is not the main theme of this article and therefore further discussions on this subject will be left for another time. 
However, the modification of the inner product due to the inclusion of the Tor terms presented above, amounts to a global boundary condition of the type introduced in [27]. 
As shown in [26] the index of the Dirac operator can be calculated in terms of the $\eta$ invariant. However, the index of the Dirac operator can be written in a cohomological form as shown in [27]. With this, a cohomological reformulation of the Atiyah-Singer index theorem is possible and results in an identification of the $\eta$ contribution to the anomaly with a cohomological term appearing in the index theorem. Through that, the coefficient structure enters the construction of the $\eta$ invariant. Demanding invariance of our construction to changes in the coefficient structure is overall a very large symmetry. Indeed, as shown previously, the coefficient structure can be of many types, including elliptic curves, various groups, Lie groups, the real numbers, the rational numbers, p-adic numbers, etc. What will be relevant here will be to impose this invariance when moving from, say, $\mathbb{Z}$ or $\mathbb{R}$ to a periodic group like $\mathbb{Z}_{p}$. In this context the modification amounts to a contribution to $\eta$ that must restore the type of periodicity that becomes invisible due to the change in coefficients, leading to precisely the expected type of correction, namely 
\begin{equation}
\eta _{D}=\sum_{k}sign(\lambda_{k})
\end{equation}
with the eigenvalues of the Dirac operator $\lambda_{k}$. 
We consider a vector bundle $E$ over a manifold $Y$ with the self-adjoint elliptic first order operator $A:C^{\infty}(Y,E)\rightarrow C^{\infty}(Y,E)$. The elliptic operator has a discrete real spectrum of eigenvalues $\lambda$ with the respective eigenfunctions $\phi_{\lambda}$. The global boundary condition is now implemented by means of a projection operator $P$ that takes the functions of $C^{\infty}(Y,E)$ and projects them on the space spanned by the eigenfunctions corresponding to positive or null eigenvalues, $\phi_{\lambda}$, $\lambda\geq 0$. We can construct the operators 
\begin{equation}
\begin{array}{c}
D=\frac{\partial}{\partial u} + A\\
\\
D^{*}=-\frac{\partial}{\partial u}+A\\
\end{array}
\end{equation}
by considering the cylindrical construction $Y\times \mathbb{R}^{+}$ of $Y$ with the half-line $u\geq 0$. As shown in [27] the projector operator implementing the global boundary condition is pseudo-differential. If we define $B=A+H$ where $H$ is the projection on the null space of $A$, then $B$ is invertible and $|B|$, the positive square root of $B^{2}$ is pseudo-differential. With this $P=\frac{1}{2}(B+|B|)$ as expected. For the operator $D$ we therefore can impose the global boundary condition 
\begin{equation}
P\cdot f(*,0)=0
\end{equation}
namely we eliminate all the positive and null eigenvalues from the spectrum. Such a condition can be re-written in the form of a inner product, as explained in [27] in the form 
\begin{equation}
\int_{Y}(f(y,0),\phi_{\lambda}(y))=0, \;\;\; \forall \lambda\geq 0
\end{equation}
For the adjoint operator the boundary condition becomes 
\begin{equation}
(1-P)\cdot f(*,0)=0
\end{equation}
It is important to make some statements about the Dirac operators taken from [28]. First, Dirac operators can be seen as a result of the quantisation of the theory of connections, and the super-trace of the heat kernel of the square of a Dirac operator is the quantisation of the Chern character of the corresponding connection. The index theorem by Atyiah and Singer is therefore a statement about the relation between the heat kernel of the square of a Dirac operator and the Chern character of the associated connection. Such a relationship holds at the level of cohomology and at the level of differential forms. 
The heat operator $e^{-t D^{2}}$ associated to a Dirac operator $D$ can be seen as an interpolation between the identity operator when $t=0$ and the projection on the kernel of the Dirac operator when $t=\infty$.
Dirac operators on a compact Riemannian manifold $M$ are defined within the context of the Clifford algebra bundle $C(M)$. The Clifford algebra at a point $x\in M$ is $C_{x}(M)$, namely the associative complex algebra generated by cotangent vectors $\alpha \in T^{*}_{x}(M)$ with its defining relation being 
\begin{equation}
\alpha_{1}\cdot \alpha_{2}+\alpha_{2}\cdot \alpha_{1}=-2(\alpha_{1},\alpha_{2})
\end{equation}
where $(\alpha_{1}, \alpha_{2})$ is the Riemannian metric on the cotangent bundle to $M$. Given an orthonormal basis $e_{i}$ of $T_{x}M$ with dual basis $e^{i}$, we can say that the Clifford algebra $C_{x}(M)$ is generated by elements $c^{i}$ satisfying 
\begin{equation}
\begin{array}{c}
(c^{i})^{2}=-1\\
\\
c^{i}c^{j}+c^{j}c^{i}=0, \forall i\neq j\\
\end{array}
\end{equation}
We can regard the Clifford algebra above as a deformation of the exterior algebra $\wedge T^{*}_{x}M$. We can define the symbol map as a bijection of the form 
\begin{equation}
\sigma_{x}(c^{i_{1}}...c^{i_{j}})=e^{i_{1}}\wedge... \wedge e^{i_{j}}
\end{equation}
The inverse of this map we can call 
\begin{equation}
c_{x}:\wedge T^{*}_{x}M\rightarrow C_{x}(M)
\end{equation}
If we have a complex $\mathbb{Z}_{2}$ graded complex bundle on $M$, let's call it $\mathcal{E}$, resulting in the splitting 
\begin{equation}
\mathcal{E}=\mathcal{E}^{+}\oplus \mathcal{E}^{-}
\end{equation}
we call $\mathcal{E}$ a bundle of Clifford modules if there is a bundle map $c:T^{*}M\rightarrow End(\mathcal{E})$ such that 
\begin{equation}
\begin{array}{c}
c(\alpha_{1})c(\alpha_{2})+c(\alpha_{2})c(\alpha_{1})=-2(\alpha_{1},\alpha_{2})\\
c(\alpha):\mathcal{E}^{+}\leftrightarrow \mathcal{E}^{-}\\
\end{array}
\end{equation}
This means that $\mathcal{E}_{x}$ is a $\mathbb{Z}_{2}$ graded module for the algebra $C_{x}(M)$. If $M$ is even dimensional the Clifford algebra is simple and we have a decomposition
\begin{equation}
End(\mathcal{E})\cong C(M)\otimes End_{C(M)}(\mathcal{E})
\end{equation}
If we consider an even-dimensional case, and $M$ is a spin manifold, there is a Clifford module $S$ called spinor bundle such that $End(S)\cong C(M)$. On such a manifold, any Clifford module can be written as a twisted spinor bundle $\mathcal{W}\otimes S$ with $\mathcal{W}=Hom_{C(M)}(S,\mathcal{E})$. Then we can define $\Gamma_{M}\in \Gamma(M,\mathbb{C}(M))$ as the chirality operator in $C(M)$ given by 
\begin{equation}
\Gamma_{M}=i^{\dim(M)/2}c^{1}...c^{n}
\end{equation}
so that $\Gamma_{M}^{2}=1$. If we regard $\mathcal{E}$ as a vector bundle on $M$ and take $\Gamma(M,\mathcal{E})$ the space of smooth sections of $\mathcal{E}$ we consider 
\begin{equation}
\mathcal{A}(M,\mathcal{E})=\Gamma(M,\wedge T^{*}M\otimes \mathcal{E})
\end{equation}
the space of differential forms on $M$ with values in $\mathcal{E}$. If $\mathcal{E}$ is a Clifford module then by the symbol map, the space of sections is isomorphic to the space of bundle valued differential forms
\begin{equation}
\Gamma(M,End(\mathcal{E}))\cong \mathcal{A}(M, End_{C(M)}(\mathcal{E}))
\end{equation}
Therefore a section $k\in \Gamma(M,End(\mathcal{E}))$ corresponds to a differential form $\sigma(k)$ with values in $End_{C(M)}(\mathcal{E})$. Now, if $M$ is a spin manifold and $\mathcal{E}=\mathcal{W}\otimes S$ is a twisted spinor bundle, then $\sigma(k)$ is a differential form with values in $End(\mathcal{W})$. A Clifford connection on a Clifford module $\mathcal{E}$ is a connection $\nabla^{\mathcal{E}}$ on $\mathcal{E}$ satisfying the formula
\begin{equation}
[\nabla_{X}^{\mathcal{E}},c(\alpha)]=c(\nabla_{X}\alpha)
\end{equation}
where $\alpha$ is a 1-form on $M$, $X$ is a vector field, and $\nabla_{X}\alpha$ is the Levi Civita derivative of $\alpha$. Then the Dirac operator is associated to the Clifford connection $\nabla^{\mathcal{E}}$ by the composition of arrows
\begin{equation}
\Gamma(M,\mathcal{E})\xrightarrow{\nabla^{\mathcal{E}}}\Gamma(M, T^{*}M\otimes \mathcal{E})\xrightarrow \Gamma(M,\mathcal{E})
\end{equation}
With respect to a local frame $e^{i}$ of $T^{*}M$, $D$ can be written as
\begin{equation}
D=\sum_{i}c^{i}\nabla_{e_{i}}^{\mathcal{E}}
\end{equation}
The heat kernel of the square of the Dirac operator is then 
\begin{equation}
\Bracket{x|e^{-tD^{2}}|y}\in Hom(\mathcal{E}_{y},\mathcal{E}_{x})
\end{equation}
and it can also be written as 
\begin{equation}
(e^{-tD^{2}}s)(x)=\int_{M}\Bracket{x|e^{-tD^{2}}|y}s(y)|dy|,\;\; \forall s\in \Gamma(M,\mathcal{E})
\end{equation}
where $|dy|$ is the Riemannian measure on $M$.
There exists an asymptotic expansion for $\Bracket{x|e^{-tD^{2}}|y}$ for small $t$ with $dim(M)=n=2l$ of the form 
\begin{equation}
\Bracket{x|e^{-tD^{2}}|y}\sim (4\pi t)^{-l}e^{-d(x,y)^{2}/4t}\sum_{i=0}^{\infty}t^{i}f_{i}(x,y)
\end{equation}
with $f_{i}$ a sequence of smooth kernels for the bundle $\mathcal{E}$ given by local functions of the curvature of $\nabla^{\mathcal{E}}$ and Riemannian curvature of $M$. $d(x,y)$ is the geodesic distance between $x$ and $y$. 
The small time differential form obtained as the image of the symbol map 
\begin{equation}
\sigma(\Bracket{x|e^{-tD^{2}}|x})\in \mathcal{A}(M, End_{C(M)}(\mathcal{E}))
\end{equation}
contains a series of differential forms. If $g$ is a unimodular Lie algebra then we define an analytic function on $g$ by 
\begin{equation}
j_{g}(X)=det(\frac{sinh(ad(X)/2)}{ad(X)/2})
\end{equation}
as the Jacobian of the exponential map $exp: g \rightarrow G$. The square root of the Jacobian would be defined in the neighbourhood of $0\in g$ as $j_{g}^{1/2}(0)=1$. 
With this, considering the Riemannian curvature matrix of $M$, $R\in \mathcal{A}^{2}(M,so(TM))$, we can choose a local orthonormal frame $e_{i}$ of $TM$ and consider the matrix $R$ with coefficients in the form of 2-forms as
\begin{equation}
R_{ij}=(Re_{i},e_{j})\in \mathcal{A}^{2}(M)
\end{equation}
Then we can use the matrix in the definition of the Jacobian for the transformation and obtain 
\begin{equation}
J_{g}(R)=det(\frac{sinh(R/2)}{R/2})
\end{equation}
where the matrix under the determinant is a matrix with even degree differential form coefficients. The determinant is invariant under conjugation by invertible matrices and therefore it is an element of $\mathcal{A}$ independent of the frame of $TM$. 
The zero form component of $J(R)$ is by definition $1$ and we can define the $\hat{A}$ genus of the manifold $M$ as
\begin{equation}
\hat{A}=J(R)^{-1/2}=det^{1/2}(\frac{R/2}{sinh{R/2}})\in \mathcal{A}(M)
\end{equation}
This is a closed differential form whose cohomology class is independent of the metric. If we use a Clifford module $\mathcal{E}$ on $M$ and we note the connection also by $\mathcal{E}$ and the curvature $F^{\mathcal{E}}$, then the twisting curvature $F^{\mathcal{F}/S}$ of $\mathcal{E}$ is defined by the formula
\begin{equation}
F^{\mathcal{E}/S}=F^{\mathcal{E}}-R^{\mathcal{E}}\in \mathcal{A}(M, End_{C(M)}(\mathcal{E}))
\end{equation}
where $R^{\mathcal{E}}(e_{i},e_{j})=\frac{1}{2}\sum_{k<l}(R(e_{i},e_{j})e_{k},e_{l})c^{k}c^{l}$
If $M$ is a spin manifold with spinor bundle $S$ and $\mathcal{E}=\mathcal{W}\otimes S$, $F^{\mathcal{E}/S}$ is the curvature of the bundle $\mathcal{W}$ then for $a\in \Gamma(M,C(M))\cong \Gamma(M, \wedge T{*}M)$, and noting the $k$-form component of $\sigma(a)$ as $\sigma_{k}(a)$ we have
\begin{equation}
\Bracket{x|e^{-tD^{2}}|x}\sim (4\pi t)^{-l}\sum_{i=0}^{\infty}t^{i}k_{i}(x)
\end{equation}
with coefficients $k_{i}\in \Gamma(M, C(M)\otimes End_{C(M)}(\mathcal{E}))$ where 
$\sigma_{j}(k_{i})=0$, $\forall j>2i$ and $\sigma(k)=\sum_{i=0}^{l}\sigma_{2i}(k_{i})\in \mathcal{A}(M,End_{C(M)}(\mathcal{E}))$. Then
\begin{equation}
\sigma(k)=det^{1/2}(\frac{R/2}{sinh(R/2)})exp(-F^{\mathcal{E}/S})
\end{equation}
If we define the index of $D$ to be the integer
\begin{equation}
ind(D)=dim(ker(D^{+}))-dim(ker(D^{-}))
\end{equation}
where $D^{\pm}$ is the restriction to $\Gamma(M,\mathcal{E}^{\pm})$ then by McKlean and Singer's formulas, the index is homotopy invariant. If $E$ is a $\mathbb{Z}_{2}$ graded vector space and $A\in End(E)$ then the supertrace $Str(A)$ is the trace of the operator $\Gamma A$, where $\Gamma\in End(E)$ is the chirality operator which is equal to $\pm$ on $E^{\pm}$. Then for each $t>0$ the index of $D$ is equal to 
\begin{equation}
ind(D)=Str(e^{-tD^{2}})=\int_{M}Str\Bracket{x|e^{-tD^{2}}|x}|dx|
\end{equation}
and therefore the index of $D$ is defined by means of the restriction to the diagonal of the heat kernel of $D^{2}$ at arbitrarily small times, and these diagonal terms of the heat kernel which are determined entirely in terms of local formulas in the curvature of the connection $\nabla^{\mathcal{E}}$ and the Riemann curvature of $M$. 
We therefore arrive at 
\begin{equation}
ind(D)=(4\pi)^{-n/2}\int_{M}Str(k_{n/2}(x))dx
\end{equation}
and therefore because
\begin{equation}
\sigma(k_{n/2}(x))_{[n]}\in \mathcal{A}^{n}(M,End_{C(M)}(\mathcal{E}))\cong\Gamma(M, End_{C(M)}(\mathcal{E}))
\end{equation}
we have 
\begin{equation}
\sigma(k_{n/2}(x))_{[n]}=det^{1/2}(\frac{R/2}{sinh(R/2)})exp(-F^{\mathcal{E}/S})
\end{equation}
In order to see how invariance with respect to coefficients in cohomology brings the inflow anomaly compensation, we have to look at how to understand characteristic classes and how their dependence on the coefficient structures manifest itself. In general, when we describe a G-bundle for a certain group $G$, the group itself is a defining part of the fibre bundle. The group brings in information and structure while the basis space (let us call it X) is lifted via the fibres and its non-trivial global structure. Such bundles over the basis space X, together with the principal group G, determines the bundle but its characterisation is usually non-trivial. However, it turns out that there exists a classifying space of the group $G$ called $\mathcal{B}G$ and in fact the morphisms from the basis space to $\mathcal{B}G$ completely classify the G-bundles over $X$ themselves. Moreover, if we introduce an abelian topological group $A$ we can define the cohomology of our classifying space $\mathcal{B}G$ as a morphism from the classifying space to the abelian topological group $A$. The same type of classification can be done towards the de-looping of the abelian group $A$, call it $\mathcal{B}^{n}A$. Indeed we have
\begin{equation}
H^{n}(\mathcal{B}G,A)\cong \pi_{0}(H(\mathcal{B}G,\mathcal{B}^{n}A))
\end{equation}
The characterisation of the bundle therefore can be done by means of its cohomology with coefficients in $G$ which is generally complicated, or by means of an (eventually spectral) decomposition of the cohomology resulting in the cohomology of $\mathcal{B}G$ with coefficients in a simpler group $A$ or equivalently $\mathcal{B}^{n}A$. Such a cohomology will contain equivalence classes of cocycles which describe the bundle. Therefore a characteristic class will be a cohomology equivalence class emerging from the following sequence
\begin{equation}
X\xrightarrow{c}\mathcal{B}G\xrightarrow{k}\mathcal{B}^{n}A
\end{equation}
where $c$ represents the map from $X$ to $\mathcal{B}G$ and it classifies the $G$-bundle. At the same time, the class $k[c]\in H^{n}(X,A)$ is a characteristic class of the bundle. 
Therefore the characteristic classes are obtained depending on the coefficient structure of their original cohomology.

\section{conclusion}
As a conclusion, in this article I introduce the first ideas regarding an alternative method of anomaly cancellation based on anti-anomalies introduced by changing the coefficient structure in the cohomological theories defining the integrals performed over the dual currents over D-branes. Thinking of coefficients in cohomology as of anomaly cancellation tools would provide us with alternatives to standard anomaly cancellation relying usually only on a higher-dimensional perspective.

\end{document}